\documentclass[10pt]{iopart}

\usepackage{graphicx}
\usepackage{dcolumn}
\usepackage{bm}
\usepackage{color}

\begin{document}

\title[Alfven Eigenmodes in LHD]{Analysis of Alfven Eigenmodes destabilization by energetic particles in Large Helical Device using a Landau-closure model}


\author{J. Varela}
\ead{rodriguezjv@ornl.gov}
\address{Oak Ridge National Laboratory, Oak Ridge, Tennessee 37831-8071}
\author{D. A. Spong}
\address{Oak Ridge National Laboratory, Oak Ridge, Tennessee 37831-8071}
\author{L. Garcia}
\address{Universidad Carlos III de Madrid, 28911 Leganes, Madrid, Spain}

\date{\today}

\begin{abstract}
Energetic particle populations in nuclear fusion experiments can destabilize Alfv\' en Eigenmodes through inverse Landau damping and couplings with gap modes in the shear Alfv\' en continua. We use the reduced MHD equations to describe the linear evolution of the poloidal flux and the toroidal component of the vorticity in a full 3D system, coupled with equations of density and parallel velocity moments for the energetic particles. We add the Landau damping and resonant destabilization effects by a closure relation. We apply the model to study the Alfv\' en modes stability in Large Helical Device (LHD) inward-shifted configurations, performing a parametric analysis in a range of realistic values of energetic particle $\beta$ ($\beta_{f}$), ratios of the energetic particle thermal/Alfv\' en velocities ($V_{th}/V_{A0}$), magnetic Lundquist numbers ($S$) and toroidal modes ($n$). The $n = 1$ and $n = 2$ TAE are destabilized although $n = 3$ and $n = 4$ TAE are weakly perturbed. The most unstable configurations are associated with density gradients of energetic particles in the plasma core: TAE are destabilized even for small energetic particle populations if their thermal velocity is lower than $0.4$ times the Alfv\' en velocity. The frequency range of MHD bursts measured in LHD are $50-70$ kHz for $n=1$ and $60-80$ kHz for $n=2$ TAE, consistent with the model predictions.
\end{abstract}

%
%
%
%
%


\vspace{2pc}

This manuscript has been authored by UT-Battelle, LLC under Contract No. DE-AC05- 00OR22725 with the U.S. Department of Energy. The United States Government retains and the publisher, by accepting the article for publication, acknowledges that the United States Government retains a non-exclusive, paid-up, irrevocable, world-wide license to publish or reproduce the published form of this manuscript, or allow others to do so, for United States Government purposes. The Department of Energy will provide public access to these results of federally sponsored research in accordance with the DOE Public Access Plan (http://energy.gov/downloads/doe-public-access-plan).

\maketitle

\ioptwocol

\section{Introduction \label{sec:introduction}}

The effect of energetic particle populations on plasma stability remains an open question for thermonuclear fusion experiments. The interaction of Alfv\' en Eigenmodes (AE) on the transport of fusion produced alpha particles, energetic hydrogen neutral beams or particle heated using ion cyclotron resonance heating (ICRF) is not well understood yet \cite{1,2,3}. Experiments in tokamaks as TFTR, JET and DIII-D or stellarators as LHD and W7-AS measured the excitation of AE, leading to a drop of the device performance \cite{4,5,6,7,8,9}. The resonance of energetic particles with velocities similar to the Alfv\' en velocity can destabilize the plasma driving instabilities that enhance particle losses, leading to a lower heating efficiency, more restrictive operation requirement for plasma ignition and also an enhancement of the diffusive losses.

If the mode frequency is small compared to the drift, bounce or transit frequencies of the energetic particles, the so called nonresonant limit, the interaction between background plasma and energetic particles leads to a stabilizing effect \cite{10,11}. In case of a resonance with a plasma instability, internal kinks \cite{12,13} or ballooning modes \cite{14} can be kinetically destabilized. AE modes can be also destabilized in low magnetic field operation regimes if the velocity of injected neutral beams particles or ICRF tails are similar to the Alfven velocity \cite{15,16}. 

The initial velocity of alpha particles and energetic particles from neutral beam injection (NBI) can exceed the Alfv\' en speed, therefore super-Alfv\' enic particles can destabilize AE, for example the toroidicity-induced shear Alfven modes (TAE), driven in the spectral gaps in between the shear Alfven continua \cite{17,18}. If the energetic particle velocity is similar to the TAE phase velocity, there is a transfer of free energy from the energetic particle density gradient to the destabilized AE gap mode \cite{19,20}. The consequence is an enhancement of alpha particle losses before thermalization \cite{21}, increasing the requirements for operations in self-sustained ignited plasmas, or a decrease of the NBI heating efficiency, also due to energetic particle losses \cite{22}.

Tangential NBI is used to heat LHD plasmas, injecting energetic hydrogen neutrals up to 180 keV. Several experiments were performed to analyze the destabilization of AE by NBI energetic particles, easily perturbed in configurations with low magnetic field (0.5 T) \cite{23,24}. The aim of present study is to analyze the AE destabilization by energetic particles in inward-shifted LHD configurations, comparing simulation results and experimental observations. 

A set of simulations are performed using an updated version of the FAR3D code \cite{25,26,27}, adding the moment equations of the energetic ion density and parallel velocity \cite{28,29}. This numerical model, with the appropriate Landau closure relations, solves the reduced non-linear resistive MHD equations including the linear wave-particle resonance effects required for Landau damping/growth \cite{30}. The code follows the evolution of a destabilized equilibria, calculated by the VMEC code \cite{31}. The results contained in this paper represent the first application of a Landau-closure moments method to a stellarator for the analysis of energetic particle instabilities. A primary motivation for such a model is its computational efficiency; this is due to its reduction of selected kinetic effects to a set of 3D fluid-like equations rather than the more conventional full 5D phase space approach \cite{32}. As described in \cite{30}, a methodology has been developed for calibrating such Landau-closure models against more complete kinetic models through optimization of the closure coefficients. This type of model is especially useful for rapid parameter/profile scans; these can be essential in modeling energetic particle instabilities since critical fast ion characteristics, such as the density profile often cannot directly be measured. It is also expected that such a model could be useful for stellarator design optimization, where rapidly evaluated physics target functions are required. Additionally, the Landau closure model described here is the only known non perturbative energetic particle stability model where it is feasible to do an eigenmode analysis; however, the current paper is based on an initial value approach. Finally, in comparison to particle-based models, this approach has the advantages of zero noise levels, exact implementation of boundary conditions and an improved ability to included extended mode coupling effects. This paper demonstrates the application of a basic Landau-closure technique. It includes Landau resonance couplings, but not thermal or fast ion FLR \cite{29}, Landau damping of the modes on the background ions/electrons \cite{28}, or coupling to acoustic waves \cite{30}. Methods for including these effects have been developed for the companion tokamak gyrofluid code TAEFL \cite{30}, and will be adapted to this 3D Landau fluid model as a topic for future research.​

This paper is organized as follows. The model equations, numerical scheme and equilibrium properties are described in section \ref{sec:model}. The simulation results are presented in section \ref{sec:simulation}. Finally, the conclusions of this paper are presented in section \ref{sec:conclusions}.

\section{Equations and numerical scheme \label{sec:model}}

For high-aspect ratio configurations with moderate $\beta$-values (of the order of the inverse aspect ratio), we can apply the method employed in Ref.\cite{33} for the derivation of the reduced set of equations without averaging in the toroidal angle to describe the evolution of the background plasma and fields. We get a reduced set of equations using the exact three-dimensional equilibrium. In this formulation, we can add linear helical couplings between mode components, which were not included in the formulation developed in Ref.\cite{33}. The effect of the energetic particle population is included in the formulation as moment of the kinetic equation truncated with a closure relation \cite{32}, describing the evolution of the energetic particles density ($n_{f}$) and velocity moments parallel to the magnetic field lines ($v_{||f}$). The coefficients of the closure relation are selected to match a two-pole approximation of the plasma dispersion function.     

In the derivation of the reduced equations we assume high aspect ratio, medium $\beta$ (of the order of the inverse aspect ratio $\varepsilon=a/R_0$), small variation of the fields and small resistivity. The plasma velocity and perturbation of the magnetic field are defined as
\begin{equation}
 \mathbf{v} = \sqrt{g} R_0 \nabla \zeta \times \nabla \Phi, \quad\quad\quad  \mathbf{B} = R_0 \nabla \zeta \times \nabla \psi,
\end{equation}
where $\zeta$ is the toroidal angle, $\Phi$ is a stream function proportional to the electrostatic potential, and $\psi$ is the perturbation of the poloidal flux.

The equations, in dimensionless form, are
\begin{equation}
\frac{{\partial \psi }}{{\partial t}} = \frac{\partial\Phi}{\partial\zeta} +  \rlap{-}\iota\frac{\partial\Phi}{\partial\theta} + \frac{\eta}{S} J_\zeta
\end{equation}
\begin{eqnarray} 
\frac{{\partial U}}{{\partial t}} = \frac{{S^{2} \beta _0 }}{{2\varepsilon ^2 }}\left( {\frac{1}{\rho }\frac{{\partial \sqrt g }}{{\partial \theta }}\frac{{\partial p}}{{\partial \rho }} - \frac{{\partial \sqrt g }}{{\partial \rho }}\frac{1}{\rho }\frac{{\partial p}}{{\partial \theta }}} \right) \nonumber\\
 +\frac{\partial J^\zeta}{\partial\zeta} +  \rlap{-}\iota\frac{\partial J^\zeta}{\partial\theta} + \frac{S^{2}\beta_{f}}{2\epsilon^{2}\rho} \left(\frac{\partial \sqrt g}{\partial \theta}\frac{\partial n_{f}}{\partial \rho} - \frac{\partial \sqrt g}{\partial \rho}\frac{\partial n_{f}}{\partial \theta} \right)
\end{eqnarray} 
\begin{equation}
\label{equation}
\frac{{\partial p}}{{\partial t}} =  \frac{dp_{eq}}{d \rho}\frac{1}{\rho}\frac{\partial \Phi}{\partial \theta} + \Gamma p_{eq} \left( \frac{\partial \sqrt g}{\partial \rho} \frac{1}{\rho} \frac{\partial \Phi}{\partial \theta} - \frac{1}{\rho} \frac{\partial \sqrt g}{\partial \theta} \frac{\partial \Phi}{\partial \rho}    \right)
\end{equation} 
\begin{eqnarray}
\label{nfast}
\frac{{\partial n_{f}}}{{\partial t}} = -S\frac{v_{th,f}^2}{\omega_{cy}} \left( \Omega_{dr} \frac{\partial n_{f}}{\partial \rho} + \Omega_{d\theta} \frac{1}{\rho} \frac{\partial n_{f}}{\partial \theta} + \Omega_{d\zeta} \frac{\partial n_{f}}{\partial \zeta} \right)  \nonumber\\
- \frac{n_{f_{0}} B_{0}}{\epsilon^2 (J + \rlap{-}\iota I)} \left( \frac{\partial}{\partial \zeta} + \rlap{-}\iota \frac{\partial}{\partial \theta} \right) v_{||f} \nonumber\\
-\epsilon^{2} n_{f_{0}} \left( \Omega_{dr} \frac{\partial \Phi}{\partial \rho} + \Omega_{d\theta} \frac{1}{\rho}\frac{\partial \Phi}{\partial \theta} + \Omega_{d\zeta} \frac{\partial \Phi}{\partial \zeta} \right) + \frac{n_{f0} q_{f}B_{0} a^2}{T_{f}}\Omega_{*}(\Phi) \nonumber\\
\end{eqnarray}
\begin{eqnarray}
\label{nfast}
\frac{{\partial v_{||f}}}{{\partial t}} = + S\frac{v_{th,f}^2}{\omega_{cy}} \left( \Omega_{dr} \frac{\partial v_{||f}}{\partial \rho} + \Omega_{d\theta} \frac{1}{\rho} \frac{\partial v_{||f}}{\partial \theta} + \Omega_{d\zeta} \frac{\partial v_{||f}}{\partial \zeta} \right)   \nonumber\\
- \sqrt{\frac{\pi}{2}} S v_{th,f} \frac{B_{0}}{\epsilon^2 (J + \rlap{-}\iota I)} \left| \frac{\partial}{\partial \zeta} + \rlap{-}\iota \frac{\partial}{\partial \theta} \right| v_{||f} \nonumber\\
+ S^2 \frac{v_{th,f}^2}{n_{f_{0}}} \frac{B_{0}}{\epsilon^2 (J + \rlap{-}\iota I)} \left( \frac{\partial}{\partial \zeta} + \rlap{-}\iota \frac{\partial}{\partial \theta} \right) n_{f} \nonumber\\
+ \frac{S^2 v_{th,f}^2 q_{f}B_{0} a^2}{T_{f}}\Omega_{*}(\Psi)
\nonumber\\
\end{eqnarray}
Here, $U =  \sqrt g \left[{ \nabla  \times \left( {\rho _m \sqrt g {\bf{v}}} \right) }\right]^\zeta$, where $\rho_m$ is the mass density. The $n_{f}$ is normalized to the density in the magnetic axis $n_{f_{0}}$, $\Phi$ to $a^2B_{0}/\tau_{R}$ and $\Psi$ to $a^2B_{0}$. All lengths are normalized to a generalized minor radius $a$; the resistivity to $\eta_0$ (its value at the magnetic axis); the time to the resistive time $\tau_R = a^2 \mu_0 / \eta_0$; the magnetic field to $B_0$ (the averaged value at the magnetic axis); and the pressure to its equilibrium value at the magnetic axis. The Lundquist number $S$ is the ratio of the resistive time to the Alfv\' en time $\tau_{A0} = R_0 (\mu_0 \rho_m)^{1/2} / B_0$. The $\rlap{-} \iota$ is the rotational transform, $v_{th,f} = \sqrt{T_{f}/m_{f}}$ the energetic particles thermal velocity normalized to the Alfv\' en velocity in the magnetic axis $v_{A0}$ and $\omega_{cy}$ the energetic particles cyclotron frequency. The $q_{f}$ is the charge, $T_{f}$ the temperature and $m_{f}$ the mass of the energetic particles. $\Omega$ operators are defined as:
\begin{equation}
\Omega_{d} = \frac{v_{th,f}^2 v_{A0}}{\omega_{cy} R_{0}} \left( \Omega_{dr}\frac{\partial}{\partial \rho} + \Omega_{d\theta}\frac{1}{\rho}\frac{\partial}{\partial \theta} + \Omega_{d\zeta}\frac{\partial}{\partial \zeta} \right) 
\end{equation}
\begin{equation}
\Omega_{*} = \frac{T_{f}}{q_{f}B_{0} a^2 \rho (J + \rlap{-}\iota I)} \frac{1}{n_{f_{0}}} \frac{dn_{f_{0}}}{d\rho} \left( I\frac{\partial}{\partial \zeta} - J\frac{\partial}{\partial \theta}  \right)
\end{equation}
with 
\begin{equation}
\Omega_{dr} = \frac{\sqrt{g}}{2\rho\epsilon^2  (J + \rlap{-}\iota I)} \left( I\frac{\partial}{\partial \zeta}\frac{1}{\sqrt{g}} - J\frac{\partial}{\partial \theta}\frac{1}{\sqrt{g}} \right) 
\end{equation}
\begin{equation}
\Omega_{d\theta} = \frac{\sqrt{g}}{2\epsilon^2  (J + \rlap{-}\iota I)^2} \left[ J\frac{\partial}{\partial \rho}(J + \rlap{-}\iota I)\frac{1}{\sqrt{g}} - \beta_{*}\rho (J + \rlap{-}\iota I) \frac{\partial}{\partial \zeta}\frac{1}{\sqrt{g}} \right] 
\end{equation}
\begin{equation}
\Omega_{d\zeta} = \frac{\sqrt{g}}{2\rho\epsilon^2  (J + \rlap{-}\iota I)^2} \left[ \beta_{*}\rho (J + \rlap{-}\iota I) \frac{\partial}{\partial \theta}\frac{1}{\sqrt{g}} - I\frac{\partial}{\partial \rho}(J + \rlap{-}\iota I)\frac{1}{\sqrt{g}} \right] 
\end{equation}
Here the $\Omega_{d}$ operator is constructed to model the average drift velocity of a passing particle and $\Omega_{*}$ models its diamagnetic drift frequency.

Equilibrium flux coordinates $(\rho, \theta, \zeta)$ are used. Here, $\rho$ is a generalized radial coordinate proportional to the square root of the toroidal flux function, and normalized to one at the edge. The flux coordinates used in the code are those described by Boozer \cite{34}, and $\sqrt g$ is the Jacobian of the coordinates transformation. All functions have equilibrium and perturbation components like $ A = A_{eq} + \tilde{A} $. 

The FAR3D code uses finite differences in the radial direction and Fourier expansions in the two angular variables. The numerical scheme is semi-implicit in the linear terms. The nonlinear version uses a two semi-steps method to ensure $(\Delta t)^2$ accuracy.

\subsection{Equilibrium properties}

\begin{figure*}[h!]
\centering
\includegraphics[width=0.8\textwidth]{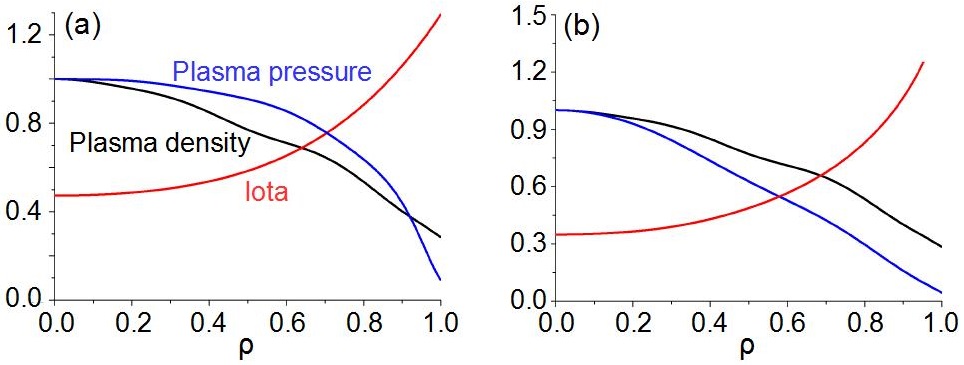}
\caption{Bulk plasma normalized pressure profile (blue line), normalized density profile (black) and rotational transform (red) for AE equil. (a) and SW equil. (b).} \label{FIG:1}
\end{figure*}

\begin{figure*}[h!]
\centering
\includegraphics[width=0.4\textwidth]{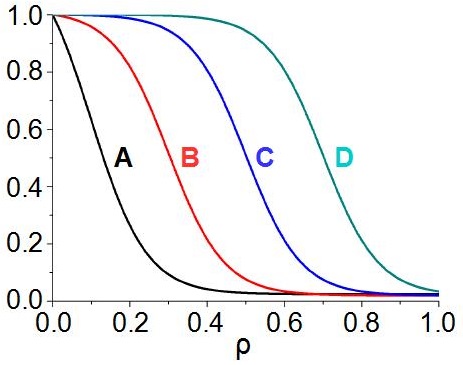}
\caption{Density profiles of the energetic particles.} \label{FIG:2}
\end{figure*}

Two fixed boundary results from the VMEC equilibrium code \cite{31} were used as input: the AE equilibria represents a LHD discharge with low magnetic field and bulk plasma density where NBI-driven Alfv\' en instabilities were observed \cite{24}, the SW equilibria is for a high magnetic field and bulk plasma density operation before the onset of a sawtooth-like event \cite{35}. The electron density and temperature profiles were reconstructed by Thomson scattering data and electron cyclotron emission. In both cases the plasma is strongly heated by neutral beams injected tangentially by three NBI lines. The vacuum magnetic axis in both equilibria is inward-shifted with $R_{\rm{axis}} = 3.76$ m for AE equil. and $3.69$ m for SW equil. The magnetic field at the magnetic axis is $0.619$ T ($2.67$ T for SW eq), the inverse aspect ratio $\varepsilon$ is $0.15$ ($0.16$ for SW eq), and $\beta_0$ is $4.2 \%$ ($1.48 \%$ for SW equil.). The injection energy of the injected energetic particles is $180$ KeV but we nominally consider only $100$ keV (energetic particle thermal velocity of $v_{th,f} = 3.1 \cdot 10^6$ m/s), resulting in an averaged Maxwellian energy equal to the average energy of a slowing-down distribution with 180 keV. The bulk electron density at the magnetic axis is $n_{e}(0) = 0.88 \cdot 10^{19}$ m$^{-3}$ ($1.6 \cdot 10^{20}$ m$^{-3}$ in SW equil.). The equilibrium rotational transform, bulk plasma normalized pressure and density profiles are plotted in figure~\ref{FIG:1}A for AE equil. and in B for SW equil. The Alfv\' en velocity in the magnetic axis is $4.55 \cdot 10^{6}$ m/s ($4.60 \cdot 10^{6}$ m/s for SW equil.) so the Alfv\' en time is $\tau_{A0} = 8.26 \cdot 10^{-7}$ s ($8.02 \cdot 10^{-7}$ s for SW equil.). The energetic particles velocity profile is considered constant for simplicity although we test several density profiles, see figure~\ref{FIG:2}, varying the location of the density gradient: near the magnetic axis (case A), the inner plasma core (case B), middle plasma (case C) and near the plasma periphery (case D). We use a constant velocity profile to emphasize the resonance efficiency between AE and bulk plasma, described by the ratio between energetic particle thermal velocity and Alfven velocity. Such simplification leads to a more didactic analysis without any loss of physical content by the model.

\subsection{Simulations parameters}

The simulations are performed with a uniform radial grid of 1000 points. The dynamic and equilibrium toroidal modes as well as the poloidal components included in the study are summarized in table~\ref{Table:1}. The kinetic closure moment equations (5) and (6) break the usual MHD parities. This is taken into account by including both parities $\sin(m\theta + n\zeta)$ and $\cos(m\theta + n\zeta)$ for all dynamic variables. The convention of the code is, in case of the pressure eigenfunction, that $n > 0$ corresponds to the Fourier component $\cos(m\theta + n\zeta)$ and $n < 0$ to $\sin(-m\theta - n\zeta)$. For example, the Fourier component for mode $1/-2$ is $\cos(-2\theta + \zeta)$ and for the mode $-1/2$ is $\sin(-2\theta + \zeta)$.  This differs from the convention used in related codes that calculate Alfv\' en continua and eigenfunctions \cite{36} since these later codes are based on ideal MHD (no kinetic closures) and no symmetry-breaking terms are present. As a result, the above modifications in the mode designations needed to accommodate symmetry breaking effects are not needed in these ideal MHD codes. The range of poloidal components covers all the possible modes along the normalized minor radius for each toroidal family, between $\rlap{-}\iota = [0.3 - 1.5]$. In the following the modes are named as $n/m$, consistent with $\rlap{-}\iota$ definition.

\begin{table}[h]
\centering
\begin{tabular}{c | c }
Dyn. toroidal mode (n) & Poloidal mode (m)  \\ \hline
$1$ & $[-8,-1]$ \\
$2$ & $[-12,-1]$ \\
$3$ & $[-16,-2]$ \\
$4$ & $[-20,-2]$ \\
Equil. toroidal mode (n) & Poloidal mode (m)  \\ \hline
$0$ & $[0,4]$ \\
$10$ & $[-7,3]$ \\
$20$ & $[-5,-1]$ \\
\end{tabular}
\caption{Dynamic and equilibrium toroidal and poloidal modes ($\rlap{-}\iota$ is defined negative in the equilibria)} \label{Table:1}
\end{table}

The range of magnetic Lundquist numbers included is $S= [10^5 - 5\cdot 10^6]$. $S = 5\cdot 10^6$ is a good approximation of the experimental value in the middle of the plasma. For lower $S$ values the plasma resistivity in the simulation is larger than in the experiment, describing a colder plasma for a early stage of the discharge.

The density ratio between energetic particles and bulk plasma ($n_{f}(0)/n_{e}(0)$) at the magnetic axis is controlled through the $\beta_{f}$ value. The ratio between energetic particle thermal velocity and Alfv\' en velocity in the magnetic axis ($v_{th,f}/v_{A0}$), controls the efficiency of the resonance coupling between AE and energetic particles.

The range of $\beta_{f}$ values included in the parametric studies goes from $0.005$ to $0.02$ and the velocity ratios between 0.1 to 1.0. The cyclotron frequency is fixed to $\omega_{cy} = 50$ (normalized to the Alfv\' en time). The energetic particle density profile used for the parametric studies is the case C (density gradient located in the middle plasma), except in the section where the effect of the energetic particle density profile is analyzed. 

In the last section we apply this model using a set of parameters obtained from LHD measurements: velocity ratio $v_{th,f}/v_{A0} = 0.68$, $\omega_{cy} = 48.98$ and $\beta_{f} = [0.01,0.03]$ for different energetic particle density profiles. In addition we analyze a case which is under development as a target for code benchmarking, based on a centrally flattened $n_{f}$ profile and $\beta_{f}$ values in the range of $[0.03,0.06]$.

\section{Simulation results \label{sec:simulation}}

The analysis is divided in textcolor{red}{two} sections. First we study the AE mode stability for the SW equilibria (high magnetic field and bulk plasma density), analyzing for $n=1$ to $n=4$ toroidal modes the growth rate (GR) and frequency (FR) in a range of $\beta_{f}$, $S$ and $v_{th,f}/v_{A0}$ parameters (namely the density of the energetic particle, plasma resistivity and resonance efficiency). In the following, we will plot the frequency in units of kHz and the growth rate in units normalized to $V_{A0}/R_{0}$.

In the second part we analyze the AE stability of the AE equil. (low magnetic field and bulk plasma density). We calculate the optimal $v_{th,f}/v_{A0}$ ratio for AE destabilization of the toroidal modes $n = 1$ and $n=2$, scanning a range of $\beta_{f}$ values in simulations whose $v_{th,f}/v_{A0}$ ratio leads to the largest GR and FR. The aim of the study is to understand the plasma equilibria properties before and after the AE destabilization. textcolor{red}{We also demostrate} the effect of the density gradient location of the energetic particles in AE stability, modifying the optimal $v_{th,f}/v_{A0}$ ratio of the resonance and the critical $\beta_{f}$. textcolor{red}{Finally} we perform a set of simulations mimicking LHD operations optimized for the destabilization of AE modes by NBI (AE equil.), comparing the model output with experimental measurements \cite{24}.

\begin{figure*}[h!]
\centering
\includegraphics[width=0.8\textwidth]{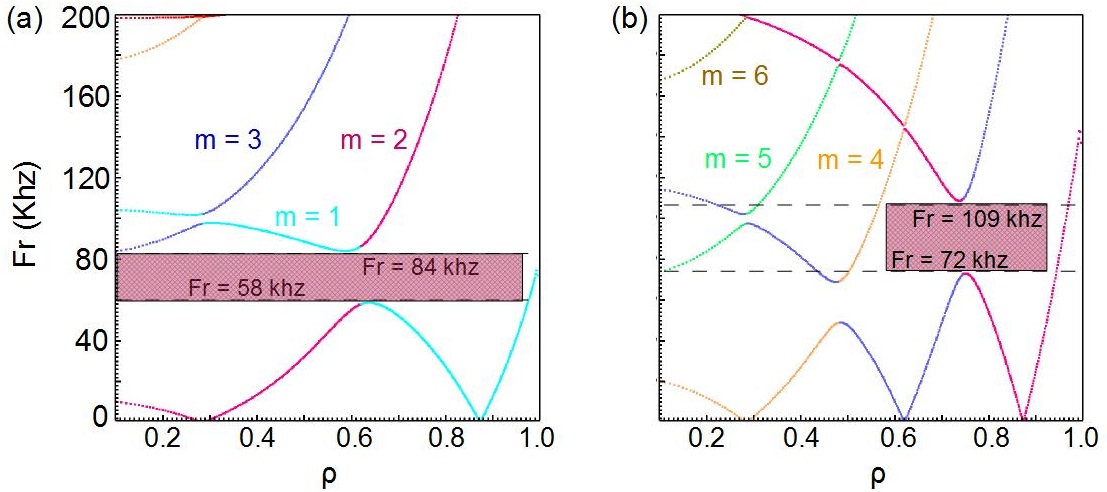}
\caption{Alfv\' en gaps of $n=1$ (a) and $n=2$ (b) toroidal modes for AE equil.}\label{FIG:3}
\end{figure*}

Figure~\ref{FIG:3} shows the Alfv\' en gaps of $n=1$ (a) and $n=2$ (b) toroidal modes for AE equilibria. Instabilities in the Alfv\' en gaps can develop because the perturbations are not damped by the continuum. In the case of $n=1$ toroidal family the range of unstable frequencies is $[58,84]$ kHz; instability can be found for $n=1$ for a range of fast ion pressure gradient locations. The frequency range of $n=2$ toroidal family is larger, [72,109] kHz, although the instabilities are mainly localized in the middle-outer region of the plasma. 

\subsection{AE stability in LHD operations with high magnetic field and bulk plasma density}     

\begin{figure*}[h!]
\centering
\includegraphics[width=0.6\textwidth]{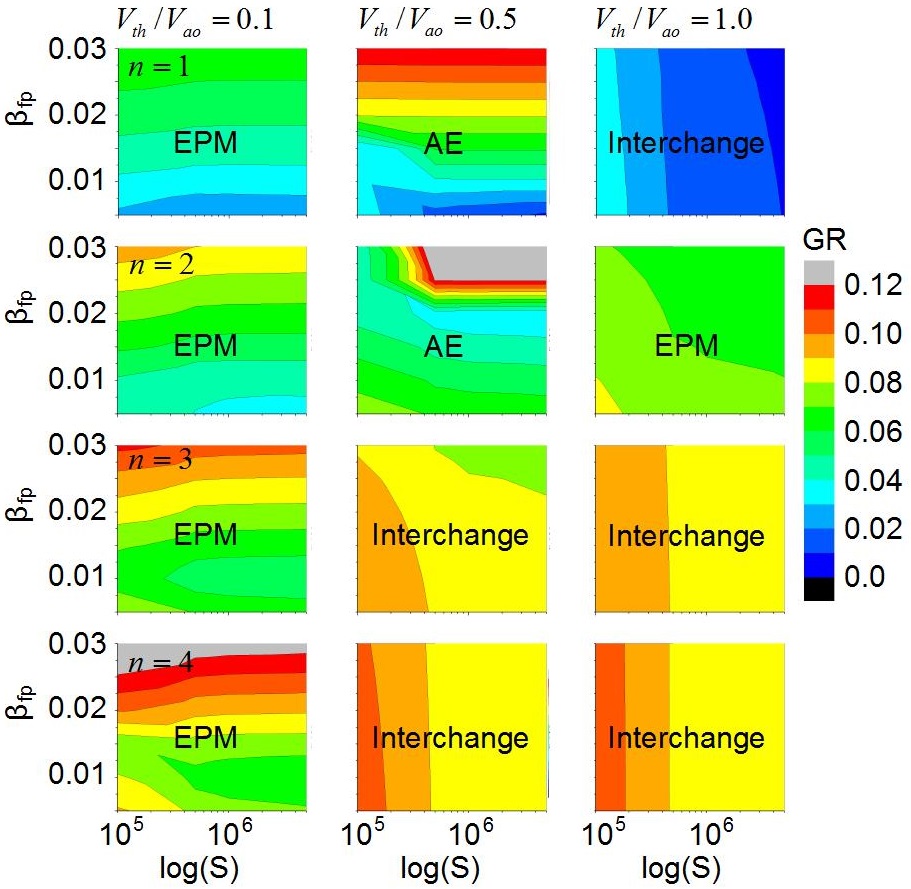}
\caption{Instability growth rate (AE, EPM or interchange mode) of the toroidal modes $n = 1$ to $4$ for different values of $\beta_{f}$ and $S$ (SW equil.). }\label{FIG:4}
\end{figure*}

Figure~\ref{FIG:4} shows with color contours the dependency of GR with $\beta_{f}$, $S$ and $v_{th,f}/v_{A0}$. The energetic particles destabilization of $n = 1$ and $n=2$ modes is weak if $v_{th,f}/v_{A0}$ is small, although it is stronger for $n = 3$ and $n=4$ modes, particularly if the $S$ value is small. The consequence is a strong destabilization of $n =3$ and $n=4$ modes at the beginning of the discharge, when the plasma is still cold, if the thermal velocity of the injected energetic particles by the NBI is around 1/10 of the Alfv\' en velocity. The NBI operation range is above that velocity ratio, at least $0.5$, so the destabilization of $n = 3$ and $4$ modes is not expected in LHD.

The $n=1$ and $n=2$ modes are strongly destabilized if $v_{th,f}/v_{A0} = 0.5$, with the GR almost independent of the $S$ values if $\beta_{f} > 0.015$. The highest GR is observed for $n = 2$ mode if $S > 5\cdot 10^5$. The $n = 3$ and $n=4$ modes show a weak dependency with $\beta_{f}$, mainly dominated by the simulation $S$ value (namely the plasma resistivity, so it is a standard MHD interchange instability), pointing out the small energetic particle destabilization. If we combine the optimal velocity ratio ($v_{th,f}/v_{A0} = 0.5$) in the NBI operation range, the weak dependency with $S$ and the strong destabilization driven for a $\beta_{f} > 0.015$, $n=1$ and $2$ modes can be destabilized even in LHD operations with high magnetic field and bulk plasma density, provided $\beta_{f}$ is sufficient.

If $v_{th,f}/v_{A0} = 1.0$ all the toroidal modes (expect $n=2$ mode) are only weakly sensitive to the energetic particle destabilization so the instabilities are MHD-like. The $n=2$ mode is unstable but the GR is less than half compared to simulations with $v_{th,f}/v_{A0} = 0.5$, therefore the unstable $n=3$ and $4$ modes dominate. In consequence, stabilization of these modes in LHD can be achieved with energetic particle populations with thermal velocities $v_{th,f} \approx v_{A0}$. After the neutral beam injection the energetic particles slow down, so it is critical to reduce the energetic particle populations with thermal velocities close to half of the Alfv\' en speed, in order to avoid the AE destabilization.

\begin{figure*}[h!]
\centering
\includegraphics[width=0.6\textwidth]{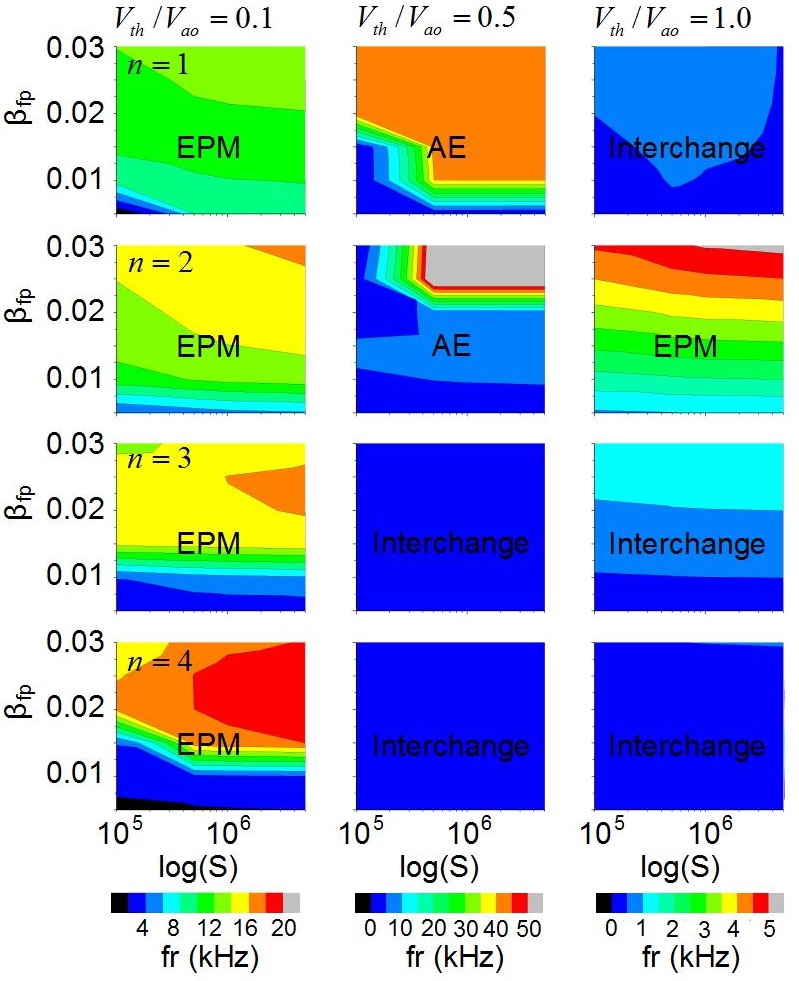}
\caption{Instability frequency (AE, EPM or interchange mode) of the toroidal modes $n = 1$ to $4$ for different values of $\beta_{f}$ and $S$ (SW equil.).}\label{FIG:5}
\end{figure*}

The instability frequency, Figure~\ref{FIG:5}, shows similar dependencies as the growth rate. The FR of $n=1$ and $n=2$ modes are small compared to $n=3$ and $n=4$ modes if $v_{th,f}/v_{A0} = 0.1$, identified as energetic particle moodes (EPM); the highest FR is observed for $n = 4$ EPM if $\beta_{f} = 0.012$ and $S = 5 \cdot 10^6$. If the velocity ratio is $v_{th,f}/v_{A0} = 0.5$, the FR of $n = 3$ and $n=4$ modes is very small compared to $n = 1$ and $n=2$ modes, identified as interchange modes and AEs respectively. The FR is independent of the $S$ value if $\beta_{f} > 0.01$ for $n = 1$ AE ($\beta_{f} > 0.02$ for $n = 2$ AE) and $S > 5\cdot 10^5$. The highest FR is reached by $n=2$ AE if $\beta_{f} = 0.03$ and $S = 5\cdot 10^6$, a $35 \%$ larger value than $n=1$ AE. Increasing the velocity ratio to $v_{th,f}/v_{A0} = 1.0$ leads to a large drop of the FR; only $n=2$ EPM shows a non negligible FR, but still 10 times smaller compared to the $v_{th,f}/v_{A0} = 0.5$ cases. In summary, not all of the instabilities correspond to AE modes: low MHD frequency modes are interchange modes that show a dependency on the $S$ parameter but not on the $\beta_{f}$. The modes with a dependency on $\beta_{f}$ but a lower FR smaller than the AEs are EPM.

\begin{figure*}[h!]
\centering
\includegraphics[width=1.0\textwidth]{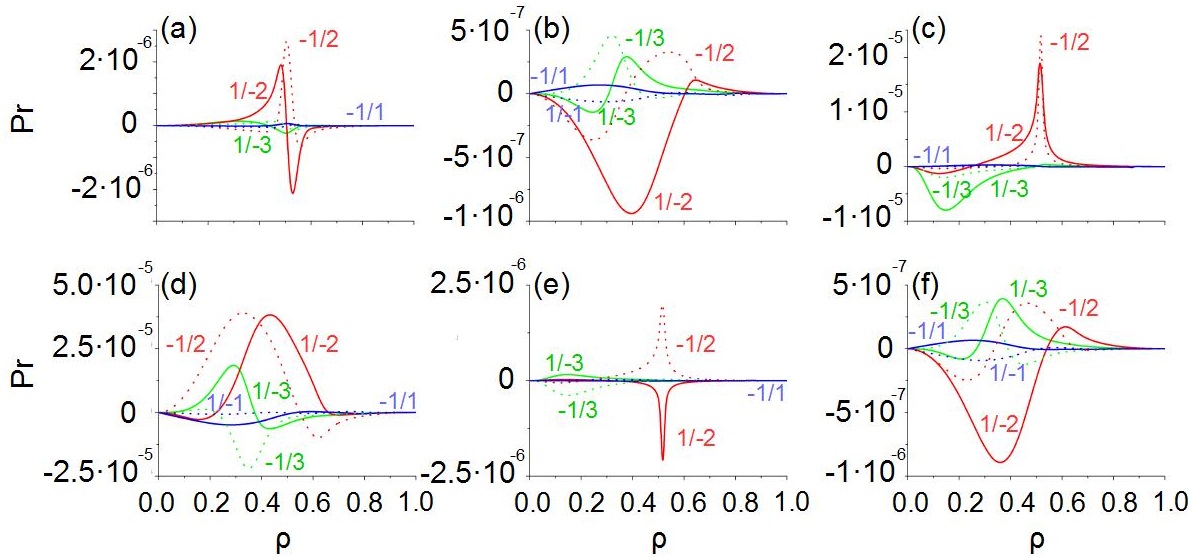}
\caption{Pressure Eigenmodes of $n=1$ mode. (a) $v_{th,f}/v_{A0} = 0.1$, $\beta_{f} = 0.02$ and $S = 5\cdot 10^6$ (b) $v_{th,f}/v_{A0} = 0.5$, $\beta_{f} = 0.02$ and $S = 5\cdot 10^6$ (c) $v_{th,f}/v_{A0} = 1.0$, $\beta_{f} = 0.02$ and $S = 5\cdot 10^6$ (d) $v_{th,f}/v_{A0} = 0.5$, $\beta_{f} = 0.02$ and $S = 1\cdot 10^5$ (e) $v_{th,f}/v_{A0} = 0.5$, $\beta_{f} = 0.005$ and $S = 5\cdot 10^6$ (f) $v_{th,f}/v_{A0} = 0.5$, $\beta_{f} = 0.03$ and $S = 5\cdot 10^6$}\label{FIG:6}
\end{figure*}

We show the pressure Eigenfunctions (Pr E.) of $n=1$ mode for different simulation parameters in Figure~\ref{FIG:6}. Recall that curves labeled with positive $n$’s correspond to $\cos(m\theta + n\zeta)$ amplitudes and those labeled with negative n’s correspond to $\sin(-m\theta - n\zeta)$ amplitudes. The upper panels, (a) to (c), analyze the effect of increasing the velocity ratio. If $v_{th,f}/v_{A0} = 0.1$, Pr E. profiles are narrow and located in the middle plasma, with the mode $1/2$ as the main driver of the instability, weakly coupled to other modes. The simulation with $v_{th,f}/v_{A0} = 0.5$ shows wider Pr E. profiles covering the inner and middle plasma, as well as the coupling between $1/2$ mode with $1/3$ and $1/1$ modes, a characteristic of toroidal AE (TAE). The simulation with $v_{th,f}/v_{A0} = 1.0$ shows the same Pr. E profile distribution than a simulation without the energetic particle destabilization, pointing out that the energetic particles don't resonante with AE and only the interchange modes are destabilized. If we compare panels (b) and (d) we observe the effect in the Pr E. profiles of reducing the $S$ parameter from $S = 5\cdot 10^6$ to $10^5$. Pr E. profiles width, distributions along the normalized minor radius and couplings between modes are similar, so the main properties of a TAE remain, indicating than the instability is weakly dependent on the plasma resistivity. We can analyze the effect of increasing $\beta_{f}$ comparing panels (e), (b) and (f). The simulation with $\beta_{f} = 0.005$, panel (E), shows narrower Pr. E profiles, weak mode coupling and modes with different parity in anti-phase, pointing out the small destabilization effect of the energetic particles. Pr E. profiles on panels (b) and (f) are similar because the $\beta_{f}$ value in both simulations is above the critical $\beta_{f}$ to trigger the AE destabilization, indicating another property of TAE: any further increase in $\beta_{f}$ above the critical value doesn't change the AE structure. The panels (a), (c) and (e) show MHD like instabilities.

\subsection{AE stability in LHD operations with low magnetic field and bulk plasma density}     

We perform a similar analysis as in the previous section for an equilibrium calculated during an experiment optimized for easier AE destabilization \cite{24}. This is based on the AE equilibria that has low magnetic field and bulk plasma density. 

\begin{figure*}[h!]
\centering
\includegraphics[width=0.6\textwidth]{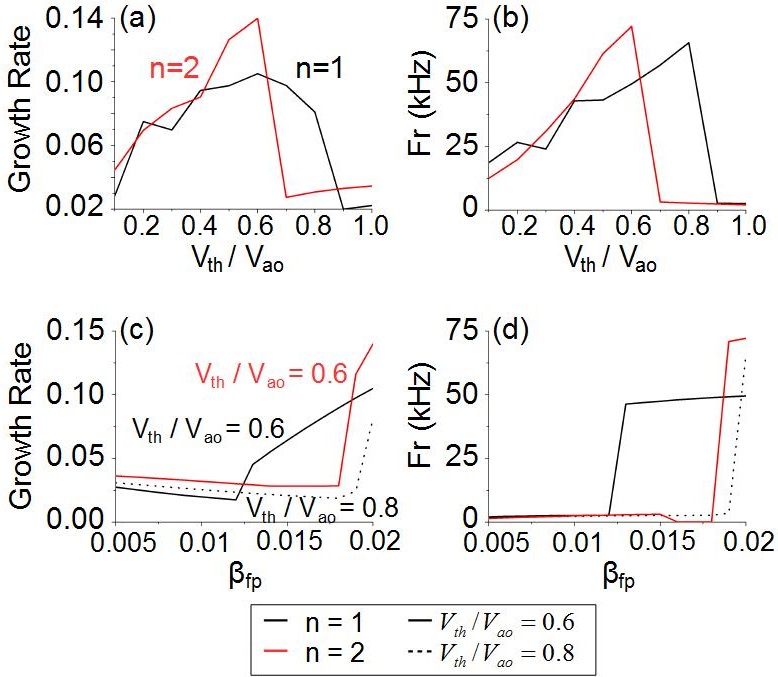}
\caption{Study of the $n=1$ (black line) and $n=2$ (red line) EP instability growth rate and frequency for different $v_{th,f}/v_{A0}$ ratios (panels a and b) and $\beta_{f}$ (panels c and d). Solid lines in panel (c and d) indicate simulations with $v_{th,f}/v_{A0} = 0.6$ and dotted lines simulations with $v_{th,f}/v_{A0} = 0.8$. (AE equil.).}\label{FIG:7}
\end{figure*}

Figure~\ref{FIG:7} shows a parametric study of the GR and FR of $n=1$ (black line) and $2$ (red line) AE for different $v_{th,f}/v_{A0}$ ratios and $\beta_{f}$ values. The largest GR is reached if $v_{th,f}/v_{A0} = 0.6$ (panel a), although the highest FR for $n=1$ AE is observed for $v_{th,f}/v_{A0} = 0.8$ (panel b). The reason is a larger destabilization of $n=1$ AE if $v_{th,f}/v_{A0} > 0.6$ compared to $n = 2$ AE, but the local maximum if $v_{th,f}/v_{A0} = 0.6$ is $40 \%$ larger for the $n = 2$ AE. These results indicate that the range of velocity ratios for an efficient resonance and destabilization of the $n=2$ AE is smaller compared to the $n=1$ AE. The AE with FR out of the FR range defined by the Alfv\' en gaps (Fig. 3) are classified as energetic particle modes (EPM), not as TAE.

We also analyze the TAE destabilization scanning the $\beta_{f}$ values range $[0.005,0.03]$, fixing $v_{th,f}/v_{A0}$ ratio to the case with the most efficient resonance, attending to the highest TAE GR and FR. For the $n=2$ TAE, $v_{th,f}/v_{A0} = 0.6$ (solid red line) leads to the highest GR and FR, although for the $n=1$ TAE, the largest GR is reached if $v_{th,f}/v_{A0} = 0.6$ (solid black line) and the highest FR if $v_{th,f}/v_{A0} = 0.8$ (dotted black line). In all cases there is a critical $\beta_{f}$ for the TAE destabilization. The critical $\beta_{f}$ is smaller for the $n=1$ TAE compared to the $n=2$ ($\beta_{f} = 0.012$ versus $0.016$) if $v_{th,f}/v_{A0} = 0.6$, although it is larger if $v_{th,f}/v_{A0} = 0.8$ ($\beta_{f} = 0.018$). The $n=2$ TAE are more unstable than $n=1$ TAE but the drive required for their destabilization is larger. In summary, the $n=1$ TAEs are easily destabilized because the critical $\beta_{f}$ is smaller and the range of $v_{th,f}/v_{A0}$ ratios for an efficient resonance is wider, so the $n=1$ TAE should be destabilized more frequently during LHD operations, although if an $n=2$ TAE is driven the instability grows faster and the FR is slightly higher compared to $n=1$ TAE.

\begin{figure*}[h!]
\centering
\includegraphics[width=1.0\textwidth]{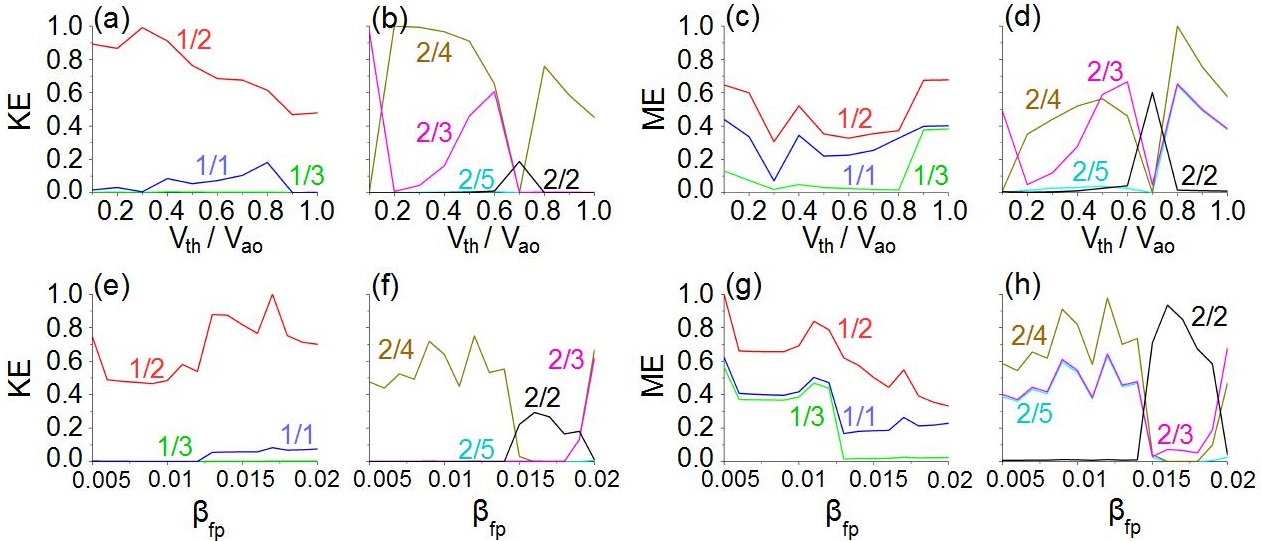}
\caption{Kinetic energy (panels a, b, e, f) and magnetic energy (panels c, d, g, h) of $n=1$ and $n=2$ EP instability dominant modes (adding the energy of both mode parities) for different $v_{th,f}/v_{A0}$ ratios and $\beta_{f}$ fixing $v_{th,f}/v_{A0} = 0.6$ (AE equil.). }\label{FIG:8}
\end{figure*}

Figure~\ref{FIG:8} shows the kinetic energy (panels a, b, e, f) and magnetic energy (panels c, d, g, h) of $n=1$ and $n=2$ AEs (adding the energy of both mode parities). The most energetic modes are plotted (namely the dominant modes) for $v_{th,f}/v_{A0} < 0.6$ are $1/2$ and $2/4$, consistent with the location of the destabilized AE in the middle plasma. For the optimal resonance case, $v_{th,f}/v_{A0} = 0.6$, there is a sharp increase of the $2/3$ mode energy, so the destabilized AE affect the middle and outer plasma ($\rho \approx 0.4-0.7$). After the velocity ratio of the resonance maxima ($v_{th,f}/v_{A0} = 0.7$), there is a drop of the $2/4$ and $2/3$ modes energy and an increase of $1/1$ and $2/2$ modes energy, indicating that the AE destabilization in the middle of the plasma is weaker but the plasma periphery is destabilized ($\rho \approx 0.9$). If $v_{th,f}/v_{A0} > 0.7$, the mode $2/4$ energy increases again and modes $1/3$ and $2/5$ in the inner plasma are destabilized, therefore the unstable AE are now located between the inner and middle plasma.

If we analyze the dominant mode energy as $\beta_{f}$ increases, the kinetic energy (KE) of the $1/2$ and $1/1$ modes grows up after the critical $\beta_{f}$ although the $2/4$ KE sharply drops while $2/2$ and $2/3$ KE increases, pointing out that $n=1$ AE is destabilized close to the middle plasma and the $n=2$ AE is destabilized closer to the plasma periphery. The modes magnetic energy (ME) near the middle plasma ($1/2$, $2/4$, $2/3$ and $2/5$) reach a local maximum for the critical $\beta_{f}$. The ME of modes close to the periphery ($1/1$ and $2/2$) in simulation above the critical $\beta_{f}$ increases while the ME of the modes near the middle plasma drops, pointing out that the source of the instability drifts from the middle to the outer plasma region. 

\begin{figure*}[h!]
\centering
\includegraphics[width=0.6\textwidth]{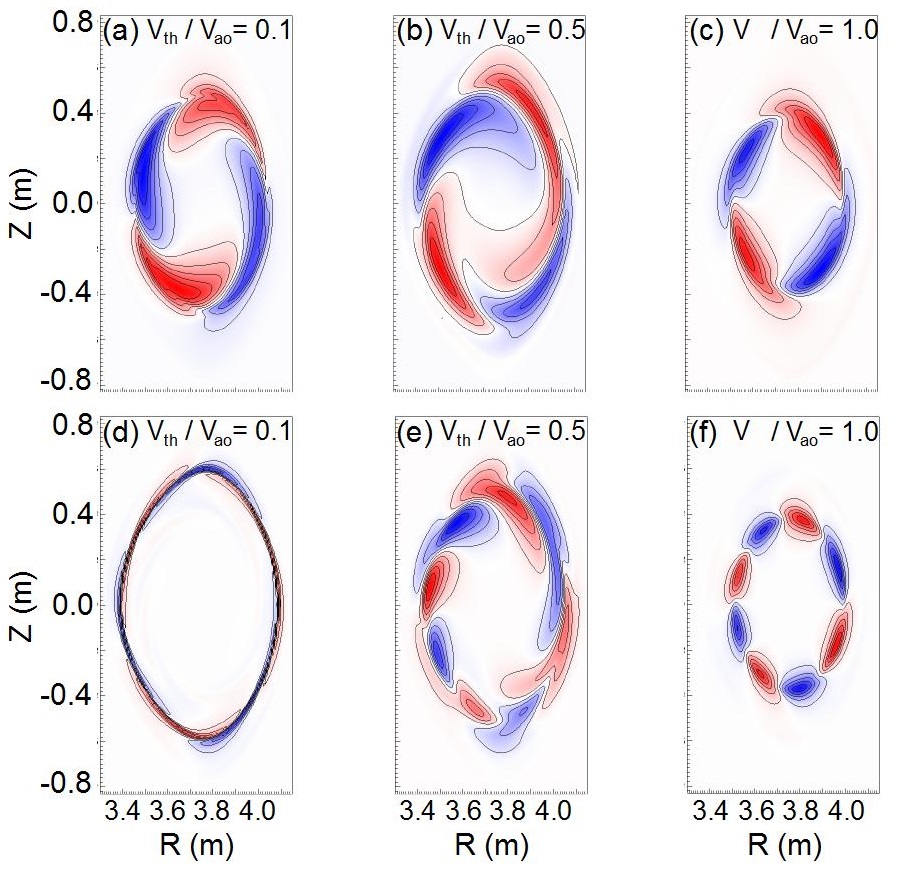}
\caption{2D plots of the $\Phi$ potential for $n=1$ (panels a to c) and $n=2$ (panels d to e) EP instability with different $v_{th,f}/v_{A0}$ ratios.}\label{FIG:9}
\end{figure*}

\begin{figure*}[h!]
\centering
\includegraphics[width=0.6\textwidth]{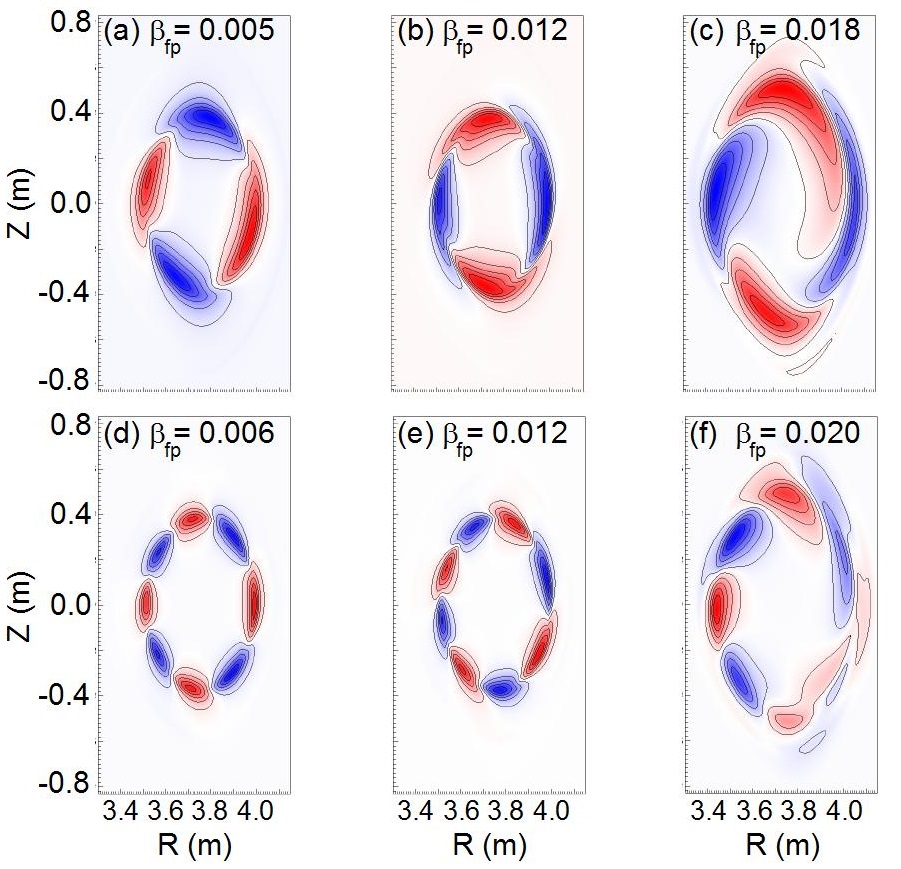}
\caption{2D plots of the $\Phi$ potential for $n=1$ (panels a to c) and $n=2$ (panels d to e) EP instability with different $\beta_{f}$ values.}\label{FIG:10}
\end{figure*}

To analyze the effect of the $v_{th,f}/v_{A0}$ ratio and $\beta_{f}$ value on the TAE structure, Figure~\ref{FIG:9} shows the $\Phi$ potential (namely the perturbation) in 2D plots for different $v_{th,f}/v_{A0}$ ratios, and in Figure~\ref{FIG:10} for different  $\beta_{f}$ values. 

If $v_{th,f}/v_{A0}$ ratio is close to the optimal resonance (Figure~\ref{FIG:9} panel b for $n=1$ AE and panel e for $n=2$ AE), the $1/2$ ($2/4$) perturbations are wide and located between the middle and outer plasma. If the $v_{th,f}/v_{A0}$ ratio is below the optimal resonance (panel a and d) the perturbations are smaller and mainly located in the middle plasma, although if $v_{th,f}/v_{A0}$ ratio is over the optimal resonance, the instability is further weakened and located closer to the inner plasma. We also observe the perturbation up-down asymmetry induced by the energetic particles destabilization. Panels d and f show a perturbation with features of an MHD-like instability, not an AE.

TAE destabilization increases if $\beta_{f}$ is higher than the critical value (Figure~\ref{FIG:10} panel c for $n=1$ TAE and panel f for $n=2$ TAE), leading to wider perturbations that cover the middle and outer plasma. Before reaching the critical $\beta_{f}$ (panels b and e), the effect of the energetic particles doesn't modify the perturbation width but drives the perturbation rotation, not observed for low values of $\beta_{f}$ (panels a and d).

The density gradient of the energetic particles is located in the middle of the plasma, which is why the TAE are destabilized mainly in the middle plasma region. To study the effect of the energetic particles density gradient on the TAE destabilization, we perform a set of simulations with different profiles of the energetic particle density (see figure~\ref{FIG:2}). We have added three new profiles of the energetic particle density: near the magnetic axis (Case A), in the inner plasma (case B) and in the plasma periphery (case D). In the next section we perform a systematic analysis of the TAE resonance attending to the $v_{th,f}/v_{A0}$ ratio and $\beta_{f}$ value.

\subsubsection{Effect of the energetic particle density profile in LHD operations with low magnetic field and bulk plasma density}    

\begin{figure*}[h!]
\centering
\includegraphics[width=0.6\textwidth]{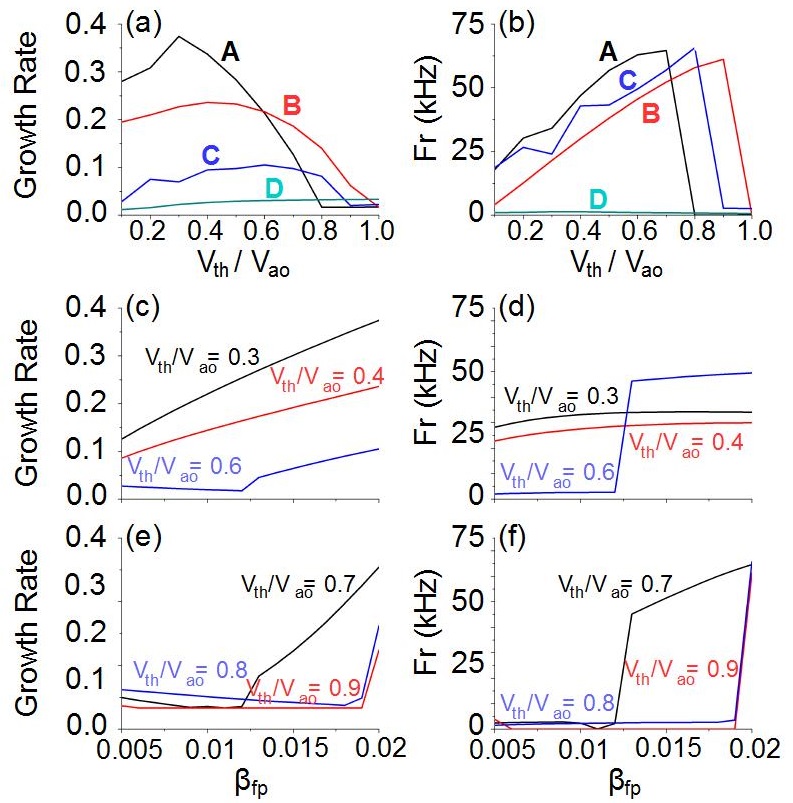}
\caption{Effect of the energetic particles density profile on the EP instability resonance efficiency, $v_{th,f}/v_{A0}$ parameter (panels a and b, the colored letter A to D indicate the energetic particle density profile case, see figure~\ref{FIG:2}. Same line color code is used on panels c to f), and critical $\beta_{f}$ value (local maximum of GR, panels c and d, local maximum of the FR, panels e and f) for $n=1$ mode.}\label{FIG:11}
\end{figure*}

\begin{figure*}[h!]
\centering
\includegraphics[width=0.6\textwidth]{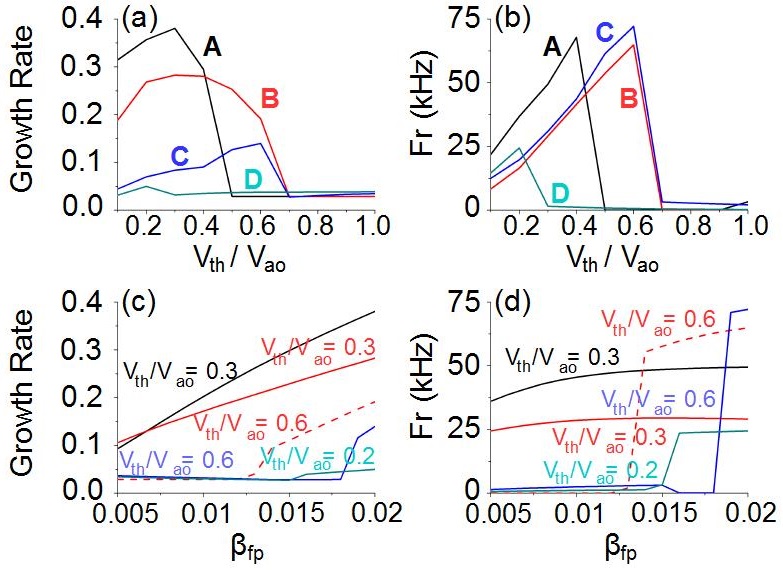}
\caption{Effect of the energetic particles density profile on the EP instability resonance efficiency (panels a and b, the colored letter A to D indicate the energetic particle density profile case, see figure~\ref{FIG:2}. Same line color code is used on panels c and d), $v_{th,f}/v_{A0}$ parameter, and critical $\beta_{f}$ value (panels c and d) for $n=2$ mode.}\label{FIG:12}
\end{figure*}

The point of maximum growth rate is displaced to lower $v_{th,f}/v_{A0}$ ratios as the density gradient is localized close to the magnetic axis (figure~\ref{FIG:11} for $n=1$ mode and figure~\ref{FIG:12} for $n=2$ mode). The GR in simulations with the case A density profile is 4 times larger as compared to case C. The maximum GR in case A takes place for $v_{th,f}/v_{A0} = 0.3$ and in case B for $v_{th,f}/v_{A0} = 0.4$. The $n=1$ instability in case D is weakly destabilized by the energetic particles and no local maximum of GR is observed, although for the $n=2$ mode there is a GR local maximum if $v_{th,f}/v_{A0} = 0.2$, almost 10 times smaller compared to case A. The local maximum of the FR is displaced to $v_{th,f}/v_{A0} > 0.6$ for the $n=1$ mode. The local maximum of the FR and GR is the same for the $n = 2$ mode, reached for $v_{th,f}/v_{A0} \leq 0.6$ (except in case B whose FR is maximum if $v_{th,f}/v_{A0} = 0.6$ and GR for $v_{th,f}/v_{A0} = 0.6$). The maximum FR is similar in all the simulations (except case D). These results predict strong EPM destabilizations if there are density gradients of energetic particles and significant populations with low thermal velocity near the magnetic axis. This scenario that can be achieved in LHD regimes with populated tails of slowing down particles with averaged velocities around $1.5 \cdot 10^{6}$ m/s (for the case of AE equil.) near the magnetic axis. Also, AE and EPM can be destabilized in LHD operations with density gradients of energetic particles in the inner plasma for large tail populations with thermal velocities around $2.0 \cdot 10^{6}$ m/s (case of AE equil.). If we consider the $v_{A0}$ radial variation effect on $v_{th,f}/v_{A0}$ (not added in present simulations) for a fixed $v_{th,f}$ value, the velocity ratio increases if the density of the bulk plasma increases or the magnetic field intensity decreases. Near the magnetic axis the bulk plasma density increases (for normal peaked profiles) and the magnetic field intensity slightly drops, so $v_{A0}$ radial variation leads to an increase of the $v_{th,f}/v_{A0}$ ratio. Consequently, for a stellarator device to have a constant $v_{th,f}/v_{A0}$ ratio, $v_{th,f}$ should be further smaller to compensate $v_{A0}$ decrease. Without such a decrease in $v_{th,f}$ AEs and EPMs GR near the magnetic axis will be smaller and FR larger. For a tokamak there is an upper limit of the bulk plasma density (Greenwald limit), therefore AE and EPM activity amelioration in the plasma core is limited.

The $\beta_{f}$ parametric study shows that, if $v_{th,f}/v_{A0} < 0.5$ and there is a peaked density profile of energetic particles near the magnetic axis (Case A) or in the inner plasma (Case B), the critical $\beta_{f}$ is lower than $0.005$ and there is a linear increase of the GR with $\beta_{f}$, several times larger compared to Case C. In consequence, such EPMs could be unstable even for small energetic particle drive (low critical $\beta_{f}$) if the density gradient of energetic particles populations is located in the inner plasma and their thermal velocities are $v_{th,f}/v_{A0} \leq 0.4$.

\begin{figure*}[h!]
\centering
\includegraphics[width=0.2\textwidth]{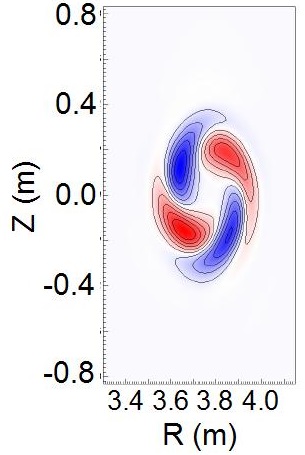}
\caption{Perturbations structure of $n=1$ EPM for the case A if $v_{th,f}/v_{A0} = 0.3$ and $\beta_{f} = 0.005$ (AE equil.). }\label{FIG:13}
\end{figure*}

Figure~\ref{FIG:13} shows the 2D mode structure for case A with $v_{th,f}/v_{A0} = 0.3$ and $28.25$ kHz. Compared to Figures~\ref{FIG:9} and~\ref{FIG:10} where the perturbations are located mainly between the middle and the outer plasma, case A simulations show wide $1/2$ perturbations between middle and inner plasma that almost reach the magnetic axis. This EPM instability can potentially reduce the LHD performance if the NBI deposition is very concentrated near the magnetic axis, although it is particularly negative for the plasma stability of fusion devices with a non negligible fraction of alpha particles in the inner plasma, because the alpha particles will be expelled from the core before thermalization so the conditions for a self sustained reaction will be more difficult to achieve.

In the next section we perform a set of simulations in the parameter range measured in LHD experiments, analyzing the optimal $v_{th,f}/v_{A0}$ ratio to destabilize AE, the effect of $\beta_{f}$ and the density profile of energetic particles. 

\subsubsection{Benchmarking with LHD measurements}     

\begin{figure*}[h!]
\centering
\includegraphics[width=0.6\textwidth]{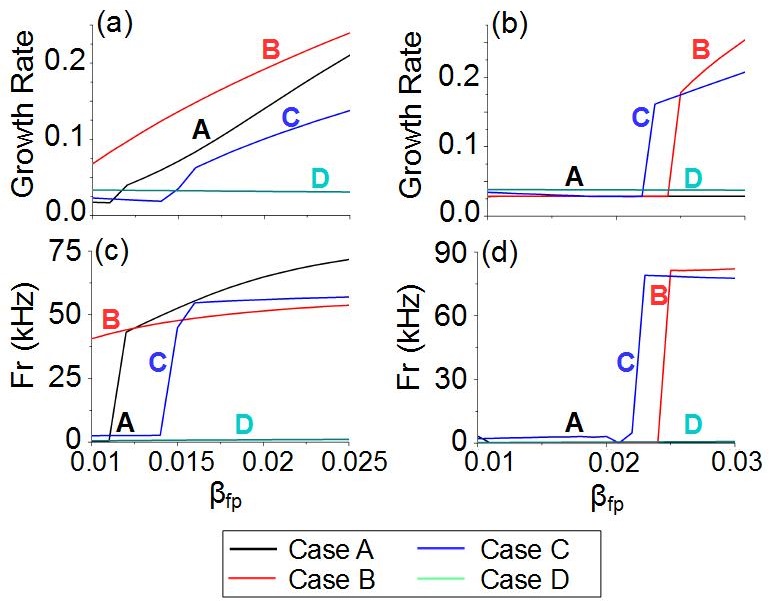}
\caption{GR and FR of $n=1$ (panels a and c) and $n=2$ (panels b and d) AE in the range of $\beta_{f}$ values measured in LHD experiments withh low magnetic field and bulk plasma density ($v_{th,f}/v_{A0} = 0.68$). Energetic particles density profiles shown by the colored letter A to D, see figure~\ref{FIG:2}.}\label{FIG:14}
\end{figure*}

Figure~\ref{FIG:14} shows the GR and FR of $n=1$ (panels a and c) and $n=2$ (panels b and d) TAE in the range of $\beta_{f}$ values measured in the experiment. Panel (a) shows that $n = 1$ TAE are unstable if the gradient of the energetic particle density is located in the middle, inner plasma or close to the magnetic axis, although it is stable if it is located near the plasma periphery. The GR increases almost linearly if $\beta_{f}$ is above the critical value, lower than $0.01$ for case A, $0.012$ case B and $0.015$ case C. The GR for case C is half compared to cases A and B. Panel (b) shows the same study for the $n=2$ TAE. The main difference compared to the $n=1$ TAE is that case A and D are stable, while the GR for cases B and C is similar. The critical $\beta_{f}$ is $0.024$ for case B and $0.022$ for case C. The FR for the $n=1$ TAE above the critical $\beta_{f}$ (panel c) is around $55$ kHz for cases B and C, almost independent of the $\beta_{f}$ value. For case A, the TAE FR increases from $45$ to $75$ kHz, a non negligible dependency with $\beta_{f}$. For $n=2$ the TAE above the critical $\beta_{f}$ (panel d), the FR dependency with $\beta_{f}$ is almost null. For cases B and C the AE FR is $85$ kHz.

LHD measurements (shot 31219) based on Mirnov coils analysis showed MHD bursts identified as $n=2$ TAE in the range of $60 - 80$ kHz, compatible with the frequencies in Fig~\ref{FIG:14}(d) \cite{24}. Also, $n=1$ TAE bursts were identified in the range of $50-70$ kHz. 

\begin{figure*}[h!]
\centering
\includegraphics[width=0.6\textwidth]{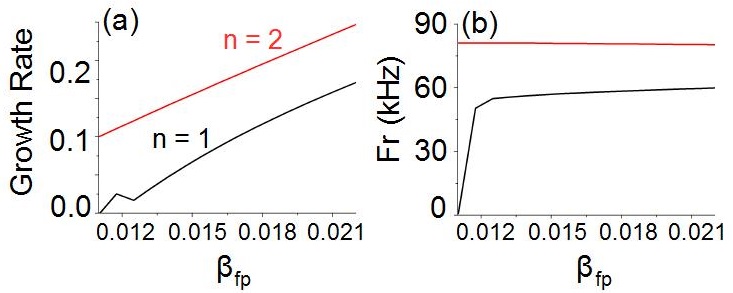}
\caption{GR (a) and FR (b) of $n=1$ and $n=2$ TAE in the range of $\beta_{f}$ values analyzed for code benchmarking, based on a centrally flattened $n_{f}$ profile ($v_{th,f}/v_{A0} = 0.68$).}\label{FIG:15}
\end{figure*}

Figure~\ref{FIG:15} shows the GR (panel a) and FR (panel b) of a test case considered for code benchmarking. In these simulations we use a                  centrally flattened $n_{f}$ profile. The GR of $n=1$ TAE is half compared to $n=2$ TAE. The critical $\beta_{f} = 0.012$ for $n=1$ TAE and lower than $0.01$ for $n=2$ TAE. The FR of the $n=1$ TAE is $80$ kHz and $60$ kHz for the $n=2$ TAE.

\section{Conclusions and discussion \label{sec:conclusions}}

This study demonstrates the viability of a Landau closure model to reproduce key features of the TAE destabilization by energetic particles. The model is expected to be a valuable tool for the analysis of TAE stability basic trends, obtained through the systematic study of realistic parametric variations.

The relatively efficient performance of the code in time and computational resources opens the possibility to analyze a large number of configurations, scanning critical parameters such as energetic particle density, ratio of thermal and Alfv\' en velocity, plasma resistivity and cyclotron frequency.

We performed a set of simulations for a realistic range of $v_{th,f}/v_{A0}$ ratios, $\beta_{f}$ values and magnetic Lundquist number, identifying $n=1$ and $n=2$ toroidal modes as the most dangerous TAE instabilities for LHD performance. LHD operations with high magnetic field and bulk plasma density lead to a weaker TAE destabilization, with frequencies almost $50 \%$ smaller compared to operations with low magnetic field and bulk plasma density, although the instability growth rate is similar. 

Analysis of the eigenfunction profiles revealed a strong coupling between dominant modes in simulations with $\beta_{f}$ above the critical value to destabilize AE, indicating that toroidal Alfv\' en Eigenmodes (TAE) are unstable. The eigenfunction profile width increases with $\beta_{f}$ until the critical $\beta_{f}$   is reached, covering wide plasma regions between the inner and outer plasma. The energetic particles destabilize TAE modes even for low $v_{th,f}/v_{A0}$ ratios, although $v_{th,f}/v_{A0}$ ratios slightly larger than the optimal resonance lead to a full AE stabilization. 

The velocity ratio $v_{th,f}/v_{A0} = 0.6$ leads to the most efficient resonance for Alfv\' en mode destabilization in LHD with low magnetic field and bulk plasma density. The $n=2$ TAE growth rate is $40 \%$ larger compared to $n=1$ TAE, although the critical $\beta_{f}$ to destabilize $n=2$ is higher, therefore $n=1$ TAE are unstable for a smaller density of the energetic particle population. If $v_{th,f}/v_{A0} \leq 0.6$, the most energetic modes are $1/2$ and $2/4$, so the instability is mainly driven in the middle plasma. Any further increase of the $v_{th,f}/v_{A0}$ ratio leads to the destabilization of $1/1$ and $2/2$ modes near the plasma periphery. If $v_{th,f}/v_{A0} \ge 0.8$ the energetic particles and TAE are no longer resonant.

Locating the energetic particle density gradient near the magnetic axis leads to the destabilization of the TAE with large growth rates, 4 to 2 times larger as compared to profiles with density gradient located in the middle plasma. If the density gradient is in the plasma periphery, no resonance between the TAE and energetic particles is observed. Configurations with the density gradient located in the inner plasma for velocity ratios $v_{th,f}/v_{A0} \ge 0.4$ are the most dangerous for LHD performance, because the critical $\beta_{f}$ is smaller than $0.005$, therefore TAE are destabilized even for small energetic particle densities, leading to wide $1/2$ perturbations extended from the middle to the inner plasma, reaching the proximity of the magnetic axis and potentially driving large losses of energetic particles in the plasma core. 

If we compare the simulation results with LHD measurements, the NBI operates in a regime above the critical $\beta_{f}$ for AE destabilization and the energetic particle thermal speed is in the ratio $v_{th,f}/v_{A0} \approx 0.6$, close to the most efficient resonance. Both $n=1$ and $n=2$ TAE are destabilized if the energetic particle density gradient is located between the middle plasma and the magnetic axis (except the $n=2$ TAE near the magnetic axis). The TAEs are stable if the density gradient is located close to the periphery. The frequency predicted by the code is consistent with MHD bursts observed in LHD operation identified as $n=1$ ($\approx 50-70$ kHz) and $n=2$ ($\approx 60-80$ kHz) TAE.

\ack
This material based on work is supported both by the U.S. Department of Energy, Office of Science, under Contract DE-AC05-00OR22725 with UT-Battelle, LLC. Research sponsored in part by the Ministerio de Economia y Competitividad of Spain under the project Nr. ENE2015-68265-P. We also want to acknowledge the LHD group in NIFS for providing us the VMEC equilibria as well as useful interactions with Y. Todo and M. Osakabe.


\newpage

\begin{thebibliography}{10}

\bibitem{1} Toi K. et al {\it Nucl. Fusion}, {\bf 44}, 217, (2004). 
\bibitem{2} Yamamoto, S. et al {\it Nucl. Fusion}, {\bf 45}, 326, (2005). 
\bibitem{3} Wilson, J. R. et al {\it Bull. Am. Phys. Soc.}, {\bf 37}, 1380, (1992).  
\bibitem{4} Wong K. L. et al {\it Phys. Rev. Lett.}, {\bf 66}, 1874, (1991). 
\bibitem{5} Sharapov, S. E. et al {\it Nucl. Fusion}, {\bf 39}, 373, (1999). 
\bibitem{6} Heidbrink, W. W. et al {\it Nucl. Fusion}, {\bf 31}, 1635, (1992).  
\bibitem{7} Duong, H. H. et al {\it Nucl. Fusion}, {\bf 33}, 749, (1993).  
\bibitem{8} Kusama, Y. et al {\it Nucl. Fusion}, {\bf 39}, 1837, (1999).  
\bibitem{9} Shinohara, K. et al {\it Nucl. Fusion}, {\bf 42}, 942, (2002).  
\bibitem{10} Rosenbluth, M. N. et al {\it Phys. Rev. Lett.}, {\bf 51}, 1967, (1983).  
\bibitem{11} White, R. B. et al {\it Phys. Rev. Lett.}, {\bf 62}, 539, (1989).  
\bibitem{12} Chen, L. et al {\it Phys. Rev. Lett.}, {\bf 52}, 1122, (1984).  
\bibitem{13} Coppi, B. et al {\it Phys. Rev. Lett.}, {\bf 57}, 2272, (1986).  
\bibitem{14} Biglari, H. et al {\it Phys. Rev. Lett.}, {\bf 67}, 3681, (1991).  
\bibitem{15} Yamamoto, S. et al {\it Phys. Rev. Lett.}, {\bf 91}, 245001, (2003).  
\bibitem{16} Darrow, D. S. et al {\it Nucl. Fusion}, {\bf 37}, 939, (1997).  
\bibitem{17} Kieras, C. et al {\it Plasma Phys.}, {\bf 28}, 395, (1982).  
\bibitem{18} Cheng, C. Z. et al {\it Phys. Fluids}, {\bf 29}, 3695, (1986).  
\bibitem{19} Fu, G. Y. et al {\it Phys. Fluids B}, {\bf 1}, 1949, (1989).  
\bibitem{20} Betti, R. et al {\it Phys. Fluids B}, {\bf 4}, 1465, (1992).  
\bibitem{21} Donne, T. et al {\it Nucl. Fusion}, {\bf 52}, 070201, (2012).  
\bibitem{22} ITER Physics Expert Group on Energetic Particles, Heating and Current Drive and ITER Physics Basis
Editors {\it Nucl. Fusion}, {\bf 39}, 2471, (1999).  
\bibitem{23} Toi, K. et al {\it Plasma Phys. and Control. Fusion}, {\bf 46}, S1, (2004).  
\bibitem{24} Osakabe, M. et al {\it Nucl. Fusion}, {\bf 46}, S911, (2006).  
\bibitem{25} Garcia, L. \textit{Proceedings of the 25th EPS International Conference, Prague, 1998}, VOL. 22A, Part II, p. 1757.
\bibitem{26} Charlton, L. A. et al {\it Journal of Comp. Physics}, {\bf 63}, 107, (1986).
\bibitem{27} Charlton, L. A. et al {\it Journal of Comp. Physics}, {\bf 86}, 270, (1990).
\bibitem{28} Spong, D. A. et al {\it Phys. Fluids B}, {\bf 4}, 3316, (1992).  
\bibitem{29} Hedrick, C. L. et al {\it Phys. Fluids B}, {\bf 4}, 3869, (1992).  
\bibitem{30} Spong, D. A. et al {\it Nucl. Fusion}, {\bf 53}, 053008, (2013).  
\bibitem{31} Hirshman, S. P. et  al {\it Phys. Fluids}, {\bf 26}, 3553, (1983).
\bibitem{32} Hammett, G. W. et  al {\it Phys. Rev. Lett.}, {\bf 64}, 3019, (1990).
\bibitem{33} Garcia, L. et al {\it Phys. Fluids B}, {\bf 2}, 2162, (1990).
\bibitem{34} Boozer, A.H. {\it Phys. Fluids}, {\bf 25}, 520, (1982).
\bibitem{35} Ohdachi, S. et al {\it Proceedings of 13th stellarator workshop 2002, Canberra}, PIIA.9, (2002).
\bibitem{36} Spong, D. A. et al. {\it Phys. Plasmas}, {\bf 17}, 022106, (2010). 

\end{thebibliography}
\end{document}